\documentclass[12pt]{article}

\def\bracket#1#2{\bigl<#1\!\bigm|\!#2\bigr>}

\usepackage{amsmath,amssymb,amscd}

\numberwithin{equation}{section}

\newcommand{\intep}{\mathbb{Z}_+}
\newcommand{\real}{\mathbb{R}}
\newcommand{\comp}{\mathbb{C}}
\newcommand{\realp}{{\mathbb{R}_+}}
\newcommand{\realm}{{\mathbb{R}_-}}
\newcommand{\realpm}{{\mathbb{R}_\pm}}
\newcommand{\realmp}{{\mathbb{R}_\mp}}
\newcommand{\compp}{{\mathbb{C}_+}}
\newcommand{\compm}{{\mathbb{C}_-}}
\newcommand{\comppm}{{\mathbb{C}_\pm}}
\newcommand{\compmp}{{\mathbb{C}_\mp}}
\newcommand{\realco}{{\mathbb{R}_x}}
\newcommand{\realmo}{{\mathbb{R}_p}}
\newcommand{\realti}{{\mathbb{R}_t}}
\newcommand{\realen}{{\mathbb{R}_E}}
\newcommand{\cF}{\mathcal{F}}
\newcommand{\cS}{\mathcal{S}} 
\newcommand{\Gammav}{\varGamma}
\newcommand{\Deltav}{\varDelta}

\begin{document}

\title{\bf Complex Eigenvalues of \\
the Parabolic Potential Barrier \\
and Gel'fand Triplet}

\author{Toshiki Shimbori$^*$ and Tsunehiro Kobayashi$^\dag$ \\
{\footnotesize\it $^*$Institute of Physics, University of Tsukuba}\\
{\footnotesize\it Ibaraki 305-8571, Japan}\\
{\footnotesize\it $^\dag$Department of General Education 
for the Hearing Impaired,}
{\footnotesize\it Tsukuba College of Technology}\\
{\footnotesize\it Ibaraki 305-0005, Japan}}

\date{}

\maketitle

\begin{abstract}
 The paper deals with the one-dimensional parabolic potential barrier 
 $V(x)={V_0-m\gamma^2 x^2/2}$, 
 as a model of an unstable system in quantum mechanics. 
 The time-independent Schr\"{o}dinger equation 
 for this model is set up as the eigenvalue problem in Gel'fand triplet 
 and its exact solutions are expressed by generalized eigenfunctions 
 belonging to complex energy eigenvalues ${V_0\mp i\Gammav_n}$ 
 whose imaginary parts are quantized as 
 ${\Gammav_n=(n+1/2)\hslash\gamma}$. 
 Under the assumption that time factors of an unstable system 
 are square integrable, 
 we provide a probabilistic interpretation of them. 
 This assumption leads to the separation 
 of the domain of the time evolution, namely all the time factors 
 belonging to the complex energy eigenvalues ${V_0-i\Gammav_n}$ 
 exist on the future part and all those 
 belonging to the complex energy eigenvalues ${V_0+i\Gammav_n}$ 
 exist on the past part. 
 In this model the physical energy distributions 
 worked out from these time factors are found to be 
 the Breit-Wigner resonance formulas. 
 The half-widths of these physical energy distributions are determined 
 by the imaginary parts of complex energy eigenvalues, 
 and hence they are also quantized. 
\end{abstract}

\thispagestyle{empty}

\pagebreak

\setcounter{page}{1}

 \section{Introduction} \label{sect.1}
 It is well known that time-dependent wave functions 
 of non-relativistic quantum mechanics satisfy 
 the time-dependent Schr\"{o}dinger equation
 \begin{equation}
  i\hslash\frac{\partial}{\partial t}\psi(t,q)
   =\hat{H}\psi(t,q), \label{1.1} 
 \end{equation}
 where $\hat{H}$ is the Hamiltonian of the system 
 expressed in the coordinate representation. 
 According to Born's probabilistic interpretation, 
 the square of modulus of 
 a time-dependent wave function $\psi(t,q)$ 
 multiplied by a small volume $\Deltav q$ 
 of the coordinate space, 
 \begin{equation}
  \left|\psi(t,q)\right|^2 \Deltav q \label{1.2} 
 \end{equation}
 is the relative probability of the particle being 
 within the small volume $\Deltav q$ of the point $q$ 
 at the time $t$. 
 If the time-dependent wave function $\psi(t,q)$ is normalized, 
 \begin{equation}
  \int\left|\psi(t,q)\right|^2 dq=1, \label{1.3} 
 \end{equation}
 then the formula \eqref{1.2} gives directly the probability. 
 In this case the expectation value of any observable $\hat{O}$ is 
 \begin{equation}
  \left<O\right>_t
   =\int\psi(t,q)^*\hat{O}\psi(t,q) dq, 
   \label{1.h}
 \end{equation}
 which is a function of $t$. 
 When a potential energy does not involve the time $t$ explicitly, 
 we can separate variables of a time-dependent wave function 
 \begin{equation}
  \psi(t,q)=T(t)u(q). \label{1.4} 
 \end{equation}
 For the state of a energy eigenvalue $E$, 
 the time factor $T(t)$ is of the form 
 \begin{equation}
  T(t)=A e^{-iEt/\hslash}, 
   \label{1.5} 
 \end{equation}
 where $A\in\comp$ is independent of $t$, 
 and equation \eqref{1.1} for it reduces to 
 the time-independent Schr\"{o}dinger equation
 \begin{equation}
  \hat{H}u(q)=Eu(q). \label{1.6} 
 \end{equation}
 In the case when the energy eigenvalue $E$ of \eqref{1.6} is 
 a real number, the formula \eqref{1.2} will reduce to 
 \begin{equation}
  \left|u(q)\right|^2 \Deltav q, \label{1.7} 
 \end{equation}
 where $u(q)$ contains $A$ as a numerical coefficient. 
 We then have, instead of \eqref{1.3} 
 \begin{equation}
  \int\left|u(q)\right|^2 dq=1, \label{1.8} 
 \end{equation}
 and instead of \eqref{1.h} we get 
 \begin{equation}
  \left<O\right>_t
   =\int u(q)^*\hat{O}u(q) dq. \label{1.p}
 \end{equation}
 Thus the probability is independent of the time. 
 We call a state belonging to the real energy eigenvalue 
 a {\it stationary state}~\cite{dirac,landau,jjs}. 
 If we cannot normalize an eigenfunction $u(q)$, we obtain 
 the expectation value of $\hat{O}$, 
 as the generalization of \eqref{1.p} 
 \begin{equation}
  \left<O\right>_t
   =\left.\int u(q)^*\hat{O}u(q) dq \right/
   \int\left| u(q)\right|^2 dq \tag{\ref{1.p}$'$}. \label{1.q} 
 \end{equation}
 
 However, the energy eigenvalue $E$ can be a complex number 
 for an unstable system in which the potential energy 
 do not have a stable stationary point. 
 The eigenfunction $u(q)$ of the Hamiltonian $\hat{H}$ 
 belonging to the complex energy eigenvalue 
 is not expressible in terms of the vector of a Hilbert space, 
 since $\hat{H}$ is a self-adjoint operator on a Hilbert space. 
 The problems of the mathematical formulations 
 of continuous spectrum or complex eigenvalues 
 has been treated by Bohm and Gadella~\cite{bohm} 
 on the basis of a {\it rigged Hilbert space} or 
 a {\it Gel'fand triplet}~\cite{gelfand}. 
 They suggested by the analysis of $S$-matrices that 
 the states belonging to continuous spectrum or complex eigenvalues 
 are expressible 
 in terms of the {\it generalized functions} of a Gel'fand triplet, 
 and their time evolution associates with a one-parameter semigroup. 

 The object of the present paper is to show that 
 the exact solution 
 of the time-independent Schr\"{o}dinger equation \eqref{1.6} 
 for an unstable potential energy becomes 
 the generalized function 
 belonging to the complex energy eigenvalue. 
 As a model Hamiltonian of the unstable system, 
 we take the one-dimensional parabolic potential barrier. 
 The collision problem of this model Hamiltonian 
 was first solved by Kemble~\cite{kemble} 
 by the method of the WKB approximation. 
 We consider the eigenvalue problem of this model Hamiltonian 
 in \S~\ref{sect.2} at one instant of time. 
 In \S~\ref{sect.2.1} this problem is worked out 
 in the coordinate representation. 
 We show that the energy eigenvalues are 
 complex numbers and the corresponding eigenfunctions 
 are expressible in terms of the generalized functions 
 of a Gel'fand triplet. 
 In \S~\ref{sect.2.2} the same problem is dealt with 
 in the momentum representation. 

 Physically these complex eigenvalues play an important part 
 in the resonance scattering. 
 Since a state of resonance is only approximately stationary, 
 its mean lifetime will be finite. 
 This means the probability of the particle being inside the system 
 tends to zero after a sufficient lapse of time, 
 and the generalized eigenfuction $u(q)$ remains finite 
 at $|q|\to\infty$. 
 This is a physical reason why 
 $u(q)$ is not expressible in a Hilbert vector 
 and does not satisfy the normalizing condition \eqref{1.8}. 
 Such a state is called a {\it non-stationary state}~\cite{dirac,jjs} 
 or a {\it quasi-stationary state}~\cite{landau}. 
 Let us now try to assume 
 the normalizing condition for the time factor 
 \begin{equation}
  \int\left|T(t)\right|^2 dt=1, \label{1.9} 
 \end{equation}
 since the mean lifetime of the non-stationary state is finite. 
 This condition will be seen later (see \S~\ref{sect.3.2}) 
 to be connected with the fact that 
 the Breit-Wigner resonance formula is normalized. 
 We can now proceed to introduce a physical interpretation 
 of the time factor $T(t)$ such that the absolute probability of 
 the particle being during the time interval $t$ to $t+\Delta t$ 
 is given by
 \begin{equation}
  \left|T(t)\right|^2 \Deltav t. \label{1.10} 
 \end{equation}
 This interpretation provides a physical description 
 of a process of time for a non-stationary state, 
 say from growth to decay. 
 When this interpretation is adapted to Born's one, 
 the square of modulus of $\psi(t,q)$ 
 corresponding to a non-stationary state, 
 multiplied by $\Delta t\Delta q$, 
 \begin{equation}
  \left|\psi(t,q)\right|^2 \Deltav t\Deltav q \label{1.11} 
 \end{equation}
 gives the relative probability of the particle being 
 within the small volume $\Delta q$ of the point $q$ 
 during the time interval $t$ to $t+\Delta t$. 
 The time average of the expectation value of $\hat{O}$ will be 
 \begin{equation}
  \left<O\right>
   =\left.\iint\psi(t,q)^*\hat{O}\psi(t,q) dt dq\right/
   \int\left| u(q)\right|^2 dq 
   \label{1.m} 
 \end{equation}
 as the generalization of \eqref{1.q}. 
 
 In \S~\ref{sect.3} the motion of the unstable system 
 consisting of the one-dimensional parabolic potential barrier 
 is considered. 
 The above theory is applied to this motion in \S~\ref{sect.3.1}. 
 In \S~\ref{sect.3.2} it is shown that the connection between 
 the physical energy distributions and 
 the complex energy eigenvalues given by \S~\ref{sect.2.1}. 
 
 \section{Eigenvalue problems of the parabolic potential barrier} 
 \label{sect.2}
  \subsection{The coordinate representation} \label{sect.2.1}
  A simple and interesting model of an unstable system in 
  quantum mechanics is 
  the one-dimensional parabolic potential barrier. 
  The Hamiltonian in quantum mechanics is 
  \begin{equation}
   \hat{H} =-\frac{\hslash^2}{2m} \frac{d^2}{dx^2}
    +V_0 -\frac{1}{2} m\gamma^2 x^2, \label{3.1}
  \end{equation}
  where $V_0\in\real$ is the maximum potential energy, 
  $m>0$ is the mass and $\gamma>0$ is 
  proportional to the square root of the curvature at $x=0$. 
  This Hamiltonian can be obtained from the Hamiltonian of 
  the harmonic oscillator \eqref{2.1} with the replacement 
  \begin{equation}
   \omega\to\mp i\gamma. \label{3.2}
  \end{equation}
  Thus $\hat{H}$ is an essentially self-adjoint on 
  a Schwartz space $\cS(\realco)$. 
  Also, two conditions are satisfied (see Appendix \ref{sect.a}). 
  \begin{enumerate}
   \item $\cS(\realco)$ is an invariant subspace of $\hat{H}$. 
	 \label{i} 
   \item $\hat{H}$ is continuous on $\cS(\realco)$. 
	 \label{ii} 
  \end{enumerate}

  The time-independent Schr\"{o}dinger equation \eqref{1.6} is
   \begin{equation}
    -\frac{\hslash^2}{2m} \frac{d^2}{dx^2}u(x)
     +\left( V_0 -\frac{1}{2} m\gamma^2 x^2\right) u(x) 
     = E u(x). \label{3.3}
   \end{equation}
   It is convenient to introduce the dimensionless variables 
   \begin{align}
    \xi\equiv&\beta x, \,\,\, 
    \beta\equiv\sqrt{\frac{m\gamma}{\hslash}}, \label{3.4}\\
    \lambda\equiv&\frac{2(E-V_0)}{\hslash\gamma}.  \label{3.5}
   \end{align}
   Equation \eqref{3.3} now becomes 
   \begin{equation}
    \frac{d^2 u}{d\xi^2}
     +\left( \lambda +\xi^2\right) u= 0. \label{3.6}
   \end{equation}
   
    \paragraph{Step 1. The asymptotic solutions \\}
    We shall now obtain the asymptotic solutions of 
    equation \eqref{3.6}. If we neglect $\lambda$ 
    in \eqref{3.6} altogether, 
    approximate solutions for large $|\xi|$ are 
    \begin{equation}
     u^\pm(\xi)\approx\xi^n e^{\pm i\xi^2/2}, \label{3.7}
    \end{equation}
    where $n$ is an arbitrary number.
    Substituting \eqref{3.7} into \eqref{3.6}, 
    we see that {\it the values $E^\pm$ of \eqref{3.5} associated with 
    $u^\pm$ are complex numbers}: 
    $$E^+\in\compm, \,\,\, E^-\in\compp, $$ 
    where the symbols $\comppm$ 
    have the meaning 
    \begin{gather*}
    \comppm\equiv\{ z=x+iy\,|\, x\in\real,\,\,\, y\in\realpm \}, \\
    \realp\equiv [0, +\infty), \,\,\, 
    \realm\equiv (-\infty, 0]. 
    \end{gather*}
    Since $\hat{H}$ is an essentially self-adjoint on $\cS(\realco)$, 
    $u^\pm$ do not belong to a Lebesgue space $L^2(\realco)$. 
    For such eigenfunctions of $\hat{H}$ 
    belonging to the complex energy eigenvalues, 
    one can use the Gel'fand triplet 
    which is formed by the Schwartz space $\cS(\realco)$ 
    as a nuclear space~\cite{bohm,bogolubov}, 
    \begin{equation}
    \cS(\realco)\subset L^2(\realco)\subset{\cS(\realco)}^\times. 
      \label{3.9}
    \end{equation}
    
    Let us treat the extension $\hat{H}^\times$ 
    of the Hamiltonian 
    to the conjugate space ${\cS(\realco)}^\times$. 
    We should be able to apply it to a generalized function 
    $u\in{\cS(\realco)}^\times$, 
    the product $\hat{H}^\times u$ 
    being defined by~\cite{bohm,bogolubov}
    $$\bracket{v}{\hat{H}^\times u}=\bracket{\hat{H} v}{u}$$
    for all functions $v\in\cS(\realco)$. 
    Taking the $x$-representatives, we get
    $$\int_{-\infty}^{\infty}v(x)^* \bigl(\hat{H}^\times u\bigr)(x)dx=
    \int_{-\infty}^{\infty}\bigl(\hat{H} v\bigr)(x)^* u(x)dx. $$
    We can transform the right-hand side 
    by partial integration and get
    $$\int_{-\infty}^{\infty}v(x)^* \bigl(\hat{H}^\times u\bigr)(x)dx=
    \int_{-\infty}^{\infty}v(x)^* \bigl(\hat{H} u\bigr)(x)dx, $$ 
    since $v$ is a rapidly decreasing function, 
    the contributions from the limits of integration vanish. 
    This gives
    $$\bracket{v}{\hat{H}^\times u}=\bracket{v}{\hat{H} u}, $$ 
    showing that 
    $$\hat{H}^\times u=\hat{H} u. $$
    Thus $\hat{H}^\times$ operating to a generalized function 
    has the meaning of $\hat{H}$ operating. 
    
    The asymptotic solutions \eqref{3.7} will then be generalized functions 
    or tempered distributions 
    in ${\cS(\realco)}^\times$. 
    We may verify that $u^\pm\in{\cS(\realco)}^\times$ by 
    the Gauss-Fresnel integral, e.g. for any 
    $x^m e^{-\alpha^2 x^2/2}\in\cS(\realco) \,\,\, 
    (m\in\intep, \alpha\in\real)$, 
    $$\int_{-\infty}^{\infty}\left( x^m e^{-\alpha^2 x^2/2}\right)^* 
    x^n e^{\pm i\beta^2 x^2/2} dx=(m+n-1)!! 
    \sqrt{\frac{2\pi}{\left( \alpha^2\mp i\beta^2\right)^{m+n+1}}}$$ 
    when $m+n$ is even and zero otherwise. 
    
    Let us examine physical properties of the asymptotic states. 
    The result of the momentum operator 
    $\hat{p}^\times =-i\hslash d/dx$ applied to 
    asymptotic solutions \eqref{3.7} is
    \begin{equation}
     -i\hslash\frac{d}{dx}u^\pm(x)\approx
      \pm\hslash\beta^2 x u^\pm(x). 
    \end{equation}
    Thus $u^+(x)$ represents particles moving outward to 
    the infinity $x=\infty$, and 
    $u^-(x)$ represents particles moving inward to 
    the origin $x=0$.
    We shall see in \S~\ref{sect.3.1} that 
    this {\it space boundary condition} is justified by 
    the asymptotic behaviors of the probability currents. 

    \paragraph{Step 2. The method of power series expansion \\}
    We shall work out the whole solutions of 
    equation \eqref{3.6} by the method of power series expansion 
    from \eqref{3.7}. 
    Put
    \begin{equation}
     u^\pm(\xi)=e^{\pm i\xi^2/2} H^\pm(\xi), 
      \label{3.14}
    \end{equation}
    introducing two new functions $H^\pm(\xi)$. 
    Equation \eqref{3.6} becomes
    \begin{equation}
     \frac{d^2H^\pm}{d\xi^2} 
      \pm 2i\xi\frac{dH^\pm}{d\xi} 
      +\left(\lambda^\pm\pm i\right)H^\pm=0. 
      \label{3.15}
    \end{equation}
    We now look for a solution of these equations in the form of 
    power series\footnote{One cannot infer that 
    $u^\pm(\xi)$ have a definite parity in this step, 
    even though the potential energy of \eqref{3.1} is 
    a symmetrical of $\xi$, since $u^\pm(\xi)$ are generally 
    degenerate~\cite{landau,jjs}.} 
    \begin{equation}
     H^\pm(\xi)=\xi^s\sum_{n=0}^{\infty} c^\pm_n\xi^n \,\,\, 
      \left(c^\pm_0\neq 0\right), \label{3.16}
    \end{equation}
    in which values of the leading order $s$ need not be integers. 
    Substituting \eqref{3.16} in \eqref{3.15} and 
    picking out coefficients of $\xi^n$, we obtain 
    \begin{equation}
     \left.
      \begin{aligned}
       s(s-1) c^\pm_0&=0, \\
       (s+1)s c^\pm_1&=0, \\
       (s+n+2)(s+n+1) c^\pm_{n+2}&=
       \left[\mp i(2s+2n+1)-\lambda^\pm\right] c^\pm_n. 
      \end{aligned} \right\} \label{3.19}
    \end{equation}
    From the indicial equations of \eqref{3.19} 
    \begin{equation}
     s=0 \,\,\,\text{or}\,\,\, 1. \label{3.20}
    \end{equation}

    We can investigate the convergence of the series \eqref{3.16} 
    on the same lines as 
    the harmonic oscillator. 
    The third of equations \eqref{3.19} give approximately, 
    when $n$ is large, 
    $$\frac{c^\pm_{n+2}}{c^\pm_n}\approx\mp i\frac{2}{n}. $$
    The series \eqref{3.16} will therefore converge like 
    $$H^\pm(\xi)\approx e^{\mp i\xi^2}$$
    for large values of $|\xi|$. Thus, from \eqref{3.14}, we have 
    $$u^\pm(\xi)\approx e^{\mp i\xi^2/2}. $$ 
    These results change the asymptotic behaviors in which 
    $u^+(\xi)$ goes off to infinity and 
    $u^-(\xi)$ arrives from infinity. 
    There are therefore in general no permissible solutions of 
    \eqref{3.3} for which the series \eqref{3.16} extend to infinity 
    on the side of large $n$. The permissible solutions, which 
    satisfy the space boundary condition, arise when 
    the series \eqref{3.16} terminate 
    on the side of large $n$. The conditions for this termination 
    of the series are 
    \begin{align}
     c^\pm_1=&0 \label{3.24} \\
     \intertext{and} 
     \lambda^\pm=&\mp i(2s+2m+1) \label{3.25} 
    \end{align}
    for some even $m$. 
    In this case we get, 
    from the recurrence formulas of \eqref{3.19}, 
    $$c^\pm_1=c^\pm_3=\cdots=c^\pm_{m-1}
    =c^\pm_{m+1}=c^\pm_{m+2}=\cdots=0.$$
    Thus $H^\pm(\xi)$ are the polynomials of degree $s+m$. 
    Since the only possible values of $s$ are $0$ or $1$, 
    we obtain from  \eqref{3.25}, by putting $s+m=n$, 
    \begin{equation}
     E^\pm_n=V_0\mp i\left(n+\frac{1}{2}\right)\hslash\gamma
      \in\compmp \,\,\, 
      \left(n=0, 1, 2, \cdots\right). \label{3.27}
    \end{equation}
    This formulas give the {\it complex energy eigenvalues} 
    for the parabolic potential barrier. 
    For generalized eigenstates of $\hat{H}$ belonging to 
    the generalized eigenvalues $E^\pm_n$, 
    the time-independent wave functions are 
    \begin{equation}
     u^\pm_n(x)=B^\pm_n e^{\pm i\beta^2x^2/2} H^\pm_n(\beta x) 
      \in{\cS(\realco)}^\times, \label{3.28}
    \end{equation}
    where $H^\pm_n(\beta x)$ are the polynomials of degree $n$ 
    (see Appendix \ref{sect.b}), and 
    $B^\pm_n$ are the numerical coefficients. 
    These numerical coefficients cannot be determined by 
    the normalizing condition \eqref{1.8}, since 
    a scalar product of generalized functions is not defined. 
    In general, a generalized function is not 
    necessarily normalized~\cite{bogolubov}. 
    However we can determine $B^\pm_n$ with the help of 
    the following argument. 
    Comparing above results with \eqref{2.7} and \eqref{2.8}, 
    we see that $E^\pm_n$ and $u^\pm_n(x)$ are the same form 
    as in the harmonic oscillator from 
    the analytical continuation \eqref{3.2} (see Appendix \ref{sect.a}).
    From \eqref{2.c}, we therefore determine that 
    \begin{equation}
     B^\pm_n=\left(\frac{\beta}{(\pm 2i)^n n! 
	      \sqrt{\pm i\pi}}\right)^\frac{1}{2} \label{3.n}
    \end{equation}
    with the correction from our definition of $H^\pm_n(\beta x)$ of 
    Appendix \ref{sect.b}. 
    
    Let us now see the properties of time-independent wave functions 
    under a space inversion. 
    Property \eqref{a.b} of Appendix \ref{sect.b} shows that 
    $u^\pm_n(x)$ are eigenstates of the parity, 
    $$u^\pm_n(-x)=(-)^n u^\pm_n(x). $$
    The validity of this result depends on 
    the space boundary condition. 
    
    The above results suggest that the eigenfunctions of 
    the parabolic potential barrier are 
    {\it non-stationary states}. 
    It is obvious physically that the observational energy spectrum of 
    this system is real and continuous. 
    But the energy eigenvalues \eqref{3.27} are complex and discrete. 
    The connection between the real continuous spectrum 
    of observed energies and 
    the complex discrete eigenvalues 
    of mathematical eigenvalue problem 
    will be dealt with in \S~\ref{sect.3.2}. 
    
  \subsection{The momentum representation} \label{sect.2.2}
  The coordinate and the momentum representatives 
  of a good function $f\in\cS(\realco)$ are connected by 
  the Fourier transformations~\cite{dirac}
  \begin{equation}
   \left.
    \begin{aligned}
     \tilde{f}(p)=&\bigl(\hat{\cF} f\bigr)(p)\equiv
     \frac{1}{\sqrt{2\pi\hslash}}
     \int_{-\infty}^{\infty}f(x) e^{-ipx/\hslash}dx, \\
     f(x)=&\bigl(\hat{\cF}^{-1} \tilde{f}\bigr)(x)\equiv
     \frac{1}{\sqrt{2\pi\hslash}}
     \int_{-\infty}^{\infty}\tilde{f}(p) e^{ipx/\hslash}dp. 
    \end{aligned} \right\} \label{3.29}
  \end{equation}
  
  Let us treat the extension $\hat{\cF}^\times$ 
  of the Fourier transformation 
  to the conjugate space ${\cS(\realco)}^\times$. 
  We should be able to apply it to a generalized function 
  ${u\in{\cS(\realco)}^\times}$, 
  the product $\hat{\cF}^\times u$ 
  being defined by~\cite{bohm,bogolubov,schwartz}
  $$\bracket{v}{\hat{\cF}^\times u}=\bracket{\hat{\cF}^{-1} v}{u}$$
  for all functions $v\in\cS(\realco)$. 
  Taking the representatives, we get
  $$\int_{-\infty}^{\infty}
  \tilde{v}(p)^* \bigl(\hat{\cF}^\times u\bigr)(p)dp=
  \int_{-\infty}^{\infty}
  \bigl(\hat{\cF}^{-1} \tilde{v}\bigr)(x)^* u(x)dx. $$
  We can transform the right-hand side 
  by the second of equations \eqref{3.29} and get
  $$\int_{-\infty}^{\infty}
  \tilde{v}(p)^* \bigl(\hat{\cF}^\times u\bigr)(p)dp=
  \int_{-\infty}^{\infty}
  \tilde{v}(p)^* \bigl(\hat{\cF} u\bigr)(p)dp, $$
  with the exchange of the order of the integrals 
  by Fubini's theorem. 
  This gives
  $$\bracket{v}{\hat{\cF}^\times u}=\bracket{v}{\hat{\cF} u}, $$ 
  showing that 
  $$\hat{\cF}^\times u=\hat{\cF} u. $$
  Thus $\hat{\cF}^\times$ operating to a generalized function 
  has the meaning of $\hat{\cF}$ operating. 

  We note that the Fourier transformation and 
  its extension applied to \eqref{3.9} give a new Gel'fand triplet 
  \begin{equation}
   \begin{CD}
    \cS(\realco) @. \,\,\,\subset\,\,\, @. 
    L^2(\realco) @. \,\,\,\subset\,\,\, @. 
    {\cS(\realco)}^\times  \\
    @V{\hat{\cF}}VV @. 
    @V{\hat{\cF}}VV @. 
    @VV{\hat{\cF}^\times}V \\
    \cS(\realmo) @. \,\,\,\subset\,\,\, @. 
    L^2(\realmo) @. \,\,\,\subset\,\,\, @. 
    {\cS(\realmo)}^\times, \label{3.30}
   \end{CD}
  \end{equation}
  which is a unitary equivalent to 
  \eqref{3.9}~\cite{bohm,bogolubov,schwartz}. 
  Written in terms of $p$-representatives, 
  the Hamiltonian \eqref{3.1} gives 
  \begin{equation}
   \hat{\cF}\hat{H}\hat{\cF}^{-1}=
    \frac{p^2}{2m}+V_0+\frac{1}{2}m\gamma^2\hslash^2\frac{d^2}{dp^2}, 
    \label{3.31}
  \end{equation}
  an essentially self-adjoint on $\cS(\realmo)$, and also 
  the two conditions (\ref{i}) and (\ref{ii}) 
  of \S~\ref{sect.2.1} are satisfied. 
  For generalized eigenstates of \eqref{3.31} belonging to 
  the generalized eigenvalues $E^\pm_n$, 
  the time-independent wave functions are 
  \begin{align}
   \tilde{u}^\pm_n(p)=&
   \tilde{B}^\pm_n e^{\mp i\tilde{\beta}^2p^2/2} 
   H^\mp_n(\tilde{\beta} p) 
   \in{\cS(\realmo)}^\times, \label{3.32}\\
   \tilde{B}^\pm_n=&(-i)^n\left(
   \frac{\tilde{\beta}}{(\mp 2i)^n n! \sqrt{\mp i\pi}}
   \right)^\frac{1}{2}, \,\,\, 
   \tilde{\beta}\equiv\frac{1}{\sqrt{m\gamma\hslash}}. 
   \label{3.32a}
  \end{align}
  Note that the phase factor $(-i)^n$ 
  appears as a result of the Fourier transformations of $H^\pm_n$. 
  
 \section{Motion in the parabolic potential barrier} 
 \label{sect.3}
  \subsection{The time evolution} \label{sect.3.1}
  Our work in \S~\ref{sect.2} has been concerned with 
  one instant of time. The present section will be devoted to 
  the variation with time of the non-stationary states. 
  
  We can get the time factors $T^\pm_n(t)$ 
  corresponding to the complex energy eigenvalues $E^\pm_n$ 
  by substituting \eqref{3.27} into \eqref{1.5}. 
  They give 
  \begin{equation}
   T^\pm_n(t)=A^\pm_n e^{-iV_0t/\hslash}e^{\mp(n+1/2)\gamma t}, 
    \label{3.34}
  \end{equation}
  where $A^\pm_n$ are the numerical coefficients. 
  The time-dependent wave functions $\psi^\pm_n(t,x)$ 
  representing a non-stationary states of 
  complex energy eigenvalues $E^\pm_n$ are, 
  according to \eqref{1.4}, 
  \begin{align}
   \psi^\pm_n(t,x)=&T^\pm_n(t) u^\pm_n(x) \notag\\
   =&A^\pm_n B^\pm_n 
   e^{-iV_0t/\hslash} 
   e^{\mp(n+1/2)\gamma t} 
   e^{\pm i\beta^2 x^2/2} 
   H^\pm_n(\beta x). \label{3.11}
  \end{align}
  
  Before proceeding to discuss the physical meanings 
  of the time factors \eqref{3.34}, 
  we shall first verify the equation of continuity. 
  From \eqref{3.11} we see now that 
  the probability densities are 
  \begin{align}
   \rho^\pm_n(t,x)
   \equiv&\left| \psi^\pm_n(t,x)\right|^2 \notag\\
   =&\left| A^\pm_n\right|^2 \left| B^\pm_n\right|^2 
   e^{\mp(2n+1)\gamma t} 
   H^\mp_n(\beta x) H^\pm_n(\beta x) \label{3.12}
  \end{align}
  and the probability currents are 
  \begin{align}
   j^\pm_n(t,x)
   \equiv&\left.\Re\left[\psi^\pm_n(t,x)^*
   \left(-i\hslash\partial\psi^\pm_n/\partial x\right)
   (t,x)\right]\right/m \notag\\ 
   =&\pm\left| A^\pm_n\right|^2 \left| B^\pm_n\right|^2 
   e^{\mp(2n+1)\gamma t} \notag\\ 
   &\times\gamma\left\{
   x H^\mp_n(\beta x) H^\pm_n(\beta x)
   \pm 2n\beta^{-1}
   \Im\left[ H^\mp_n(\beta x) H^\pm_{n-1}(\beta x)\right]\right\}. 
   \label{3.13}
  \end{align}
  Equations \eqref{3.12} and \eqref{3.13} depend on the time $t$ 
  which satisfy the equation of continuity 
  \begin{equation}
   \frac{\partial}{\partial t}\rho^\pm_n(t,x)+
    \frac{\partial}{\partial x}j^\pm_n(t,x)=0, 
    \label{3.z}
  \end{equation}
  with the help of formulas \eqref{a.4} and \eqref{a.5} 
  of Appendix \ref{sect.b}. 
  Note that the numerical coefficients $A^\pm_n$ and $B^\pm_n$ 
  do not affect the validity of \eqref{3.z}. 
  
  We must examine how the probability currents $j^\pm_n(t,x)$ 
  behave for large values of $|x|$. The second term in the 
  $\{ \}$ brackets in \eqref{3.13} can be neglected as 
  $|x|\to\infty$. We are left with 
  \begin{align*}
   j^\pm_n(t,x)\approx&
   \pm 2^{2n}\left| A^\pm_n\right|^2 \left| B^\pm_n\right|^2 
   e^{\mp(2n+1)\gamma t} 
   \gamma\beta^{2n}x^{2n+1} \\
   \approx& \pm e^{\mp(2n+1)\gamma t} x^{2n+1}. 
  \end{align*}
  These asymptotic behaviors of probability currents show that 
  the index $+$ ($\psi^+_n$, $u^+_n$, etc.) 
  means only outward moving particles and 
  the index $-$ ($\psi^-_n$, $u^-_n$, etc.) 
  means only inward moving particles. 
  These asymptotic behaviors are the one assumed in 
  \S~\ref{sect.2.1} for dealing with the space boundary condition. 
  
  Let us now examine how the probability densities $\rho^\pm_n(t,x)$ 
  vary with time. From \eqref{3.12}, $\rho^\pm_n(t,x)$ will tend to 
  infinity as $t\to\mp\infty$. The result is a necessary consequence 
  of the form of time-dependent wave functions \eqref{3.11}. 
  In any case the time-dependent wave functions $\psi^\pm_n(t,x)$ 
  must not tend to infinity as $t\to\mp\infty$, or 
  they will represent states that have no physical meaning. 
  The way of escape from the difficulty lies in the 
  domains of the time factors \eqref{3.34}. We assume that 
  $T^+_n(t)$ exists only on the future part, for which $t>0$, and 
  $T^-_n(t)$ exists only on the past part, for which $t<0$, i.e. 
  \begin{alignat*}{3}
   T^+_n(t)=&A^+_n e^{-iV_0t/\hslash}e^{-(n+1/2)\gamma t} 
   \,\,\,&\text{when}\,\,\, t>&0, \\
   T^-_n(t)=&A^-_n e^{-iV_0t/\hslash}e^{(n+1/2)\gamma t} 
   \,\,\,&\text{when}\,\,\, t<&0. 
  \end{alignat*}
  These equations can be combined into the single equations 
  \begin{equation}
   T^\pm_n(t)=A^\pm_n e^{-iV_0t/\hslash}e^{\mp(n+1/2)\gamma t} 
    \theta(\pm t), \label{3.34a}
  \end{equation}
  where $\theta(t)$ is the Heaviside step function. 
  We take the squares of the moduli of the time factors \eqref{3.34a} 
  \begin{equation}
   \left| T^\pm_n(t)\right|^2=\left| A^\pm_n\right|^2 
    e^{\mp(2n+1)\gamma t} \theta(\pm t). \label{3.34b}
  \end{equation}
  To interpret the result \eqref{3.34b}, we may suppose that 
  the index $+$ ($\psi^+_n$, $T^+_n$, etc.) 
  refer to the {\it states of decay} when $t>0$, while 
  the index $-$ ($\psi^-_n$, $T^-_n$, etc.) 
  refer to the {\it states of growth} when $t<0$.
  This condition expresses mathematically that 
  \begin{equation}
   T^+_n(t)\in L^2_+(\realti), \,\,\, 
    T^-_n(t)\in L^2_-(\realti), \label{3.ca}
  \end{equation}
  where spaces $L^2_\pm(\realti)$ are defined by 
  \begin{equation}
   L^2_\pm(\realti)\equiv\bigl\{ f\in L^2(\realti) \bigm| 
    f(t)=0 \,\,\,\mathrm{a.e.}\,\,\,  t\in\realmp \bigr\}. 
    \label{3.c}
  \end{equation}
  This condition will be referred to as the 
  {\it time boundary condition}. 
  
  These boundary conditions require that 
  $\psi^-_n(t,x)$ represents the growing state 
  lying on the past part together with only inward moving particle 
  and $\psi^+_m(t,x)$ represents the decaying state 
  lying on the future part together with only outward moving particle. 
  We must now obtain the probability of a transition 
  taking place from state $\psi^-_n$ to state $\psi^+_m$ 
  at time $t=0$. This gives us the ``selection rule'' at time $t=0$, 
  and can be worked out from the following transition matrix elements 
  \begin{equation}
   S_{mn}=\bracket{u^+_m}{u^-_n}\equiv
    \int_{-\infty}^{\infty}u^+_m(x)^* u^-_n(x) dx, 
    \label{3.e}
  \end{equation}
  the last definition expresses $\bracket{u^+_m}{u^-_n}$ 
  in terms of $x$-representatives. 
  When we substitute for $u^\pm_n(x)$ 
  their value given by \eqref{3.28}, equation \eqref{3.e} gives 
  $$S_{mn}=\beta^{-1}{B^+_m}^* B^-_n
  \int_{-\infty}^{\infty} H^+_m(\xi)^* H^-_n(\xi)
  e^{-i\xi^2} d\xi $$
  with the help of \eqref{3.4}. 
  The right-hand side becomes, from an application of 
  formula \eqref{a.6} of Appendix \ref{sect.b}, 
  \begin{equation}
   S_{mn}=\delta_{mn} \label{3.ea}
  \end{equation}
  with the help of \eqref{3.n}. 
  The transition matrix is then a unit matrix. 
  This result expresses that 
  the state $\psi^-_n\,\,\,(n=0,1,2,\cdots)$ on the past part 
  is connected with 
  the state $\psi^+_n$ having the same label $n$ on the future part. 
  Thus our selection rule at time $t=0$ is that 
  only those transitions can take place in which 
  the complex energy changes from $E^-_n$ to $E^+_n$. 
  
  Our problem now is to obtain 
  the numerical coefficients $A^\pm_n$ of \eqref{3.34a}. 
  We shall distinguish between the growing state $\psi^-_n$ 
  and the decaying state $\psi^+_n$ of the unstable system, 
  with the growth and decay processes governed independently each by its
  own probability law~\cite{dirac}. 
  We shall require that 
  {\it the total number of growing particles is equal to 
  the total number of decaying particles}, i.e. 
  \begin{equation}
   \int_{-\infty}^0 \rho^-_n(t,x) dt =
    \int_0^\infty \rho^+_n(t,x) dt. \label{3.ix} 
  \end{equation}
  By substituting for $\rho^\pm_n$ here their values given by 
  \eqref{3.12}, we now see that this condition \eqref{3.ix} is 
  independent of $x$. Thus \eqref{3.ix} provides the condition 
  for the time factor. 
  From \eqref{3.ix} we can require 
  {\it the normalizing condition for the time factor}, 
  as in equation \eqref{1.9} of \S~\ref{sect.1} 
  \begin{equation}
   \int_{-\infty}^0 \left|T^-_n(t)\right|^2 dt =
    \int_0^\infty \left|T^+_n(t)\right|^2 dt =1. \label{3.i} 
  \end{equation}
  If we now apply these conditions \eqref{3.i} 
  to the time factors \eqref{3.34a}, we have 
  \begin{equation}
   A^\pm_n=\sqrt{(2n+1)\gamma}, \label{3.k}
  \end{equation}
  where the phase factors are chosen unity.\footnote{We choose 
  these phase factors so that 
  the time-dependent wave functions are simple forms 
  under a time reversal.}
  
  Let us now apply the rules of \eqref{1.10} for the interpretation 
  of time factors. We find that the expectation values of $t$ 
  for the states $\psi^\pm_n$ are 
  \begin{align}
   \bigl< t\bigr>^\pm_n &\equiv
   \int_{-\infty}^\infty t\left|T^\pm_n(t)\right|^2 dt \notag \\
   &=\left| A^\pm_n\right|^2 
   \int_{-\infty}^\infty t e^{\mp(2n+1)\gamma t} \theta(\pm t) dt 
   \notag \\ 
   &=\pm\frac{1}{(2n+1)\gamma}, \label{3.52x}
  \end{align}
  respectively. Again, we find 
  \begin{align*}
   \bigl< t^2\bigr>^\pm_n &\equiv 
   \int_{-\infty}^\infty t^2 \left|T^\pm_n(t)\right|^2 dt \notag \\
   &=\left| A^\pm_n\right|^2 
   \int_{-\infty}^\infty t^2 e^{\mp(2n+1)\gamma t} \theta(\pm t) dt 
   \notag \\
   &=\frac{2}{[(2n+1)\gamma]^2}. 
  \end{align*}
  Hence the positive square roots of the variance 
  of $t$ for the states $\psi^\pm_n$ are 
  \begin{align}
   (\Deltav t)^\pm_n &\equiv
   \sqrt{\bigl<\bigl(t-\bigl< t\bigr>^\pm_n\bigr)^2\bigr>^\pm_n} 
   \notag \\
   &=\sqrt{\bigl< t^2\bigr>^\pm_n -\bigl(\bigl< t\bigr>^\pm_n\bigr)^2} 
   \notag \\
   &=\frac{1}{(2n+1)\gamma}. 
   \label{3.52}
  \end{align}
  Thus $(\Deltav t)^+_n$ is equal to $(\Deltav t)^-_n$, 
  since the probability densities $\left| T^+_n(t)\right|^2$ 
  and $\left| T^-_n(t)\right|^2$ are symmetrical 
  with respect to the center $t=0$. 
  We may call these quantities $ (\Deltav t)^\pm_n$ 
  the {\it mean lifetimes}. 
  Equations \eqref{3.52} inform us that 
  {\it the mean lifetimes are quantized 
  in the unstable system with the Hamiltonian \eqref{3.1}}. 
  These results are closely connected with 
  Heisenberg's uncertainty relation between 
  time and energy. This matter will be dealt with 
  in the next section. 
  
  Let us return to the time-dependent wave functions \eqref{3.11}. 
  We shall verify finally a time reversal. 
  By application of property \eqref{a.c} of Appendix \ref{sect.b}, 
  we get 
  $$\psi^\pm_n(-t,x)^*=\psi^\mp_n(t,x), $$
  with the help of \eqref{3.k} and also of equation \eqref{3.n}. 
  Similarly we have in the $p$-representation 
  $$\tilde{\psi}^\pm_n(-t,-p)^*=\tilde{\psi}^\mp_n(t,p), $$
  since the phase factor in equation \eqref{3.32a} and 
  the parity of $H^\pm_n(\tilde{\beta}p)$ cancel out. 
  We see in this way that a time reversal occurs 
  resulting in the interchange of states 
  on the future part and on the past part. 
  It is known as the {\it reciprocity theorem}~\cite{landau}. 
  
  \subsection{The energy distribution} \label{sect.3.2}
  The time factor and the {\it physical} energy dependence of 
  a function $f\in L^2_\pm(\realti)$ are connected by 
  the inverse Fourier transformations 
  \begin{equation}
   \left.
    \begin{aligned}
     \tilde{f}(E)=&\bigl(\hat{\cF}^{-1} f\bigr)(E)\equiv
     \frac{1}{\sqrt{2\pi\hslash}}
     \int_{-\infty}^{\infty}f(t) e^{iEt/\hslash}dt, \\ 
     f(t)=&\bigl(\hat{\cF} \tilde{f}\bigr)(t)\equiv
     \frac{1}{\sqrt{2\pi\hslash}}
     \int_{-\infty}^{\infty}\tilde{f}(E) e^{-iEt/\hslash}dE,
    \end{aligned} \right\} \label{3.33}
  \end{equation}
  since the theory of relativity puts 
  physical energy in the same relation to time as 
  momentum to coordinate (see equations \eqref{3.29}). 
  
  We must now evaluate 
  the physical energy dependence $\tilde{T}^\pm_n(E)$ 
  for the time factors \eqref{3.34a}, 
  and obtain the probability density of them. 
  We make use of the Paley-Wiener theorem, which states that 
  the inverse Fourier transformation is a unitary mapping 
  from $L^2_\pm(\realti)$ onto $H_\pm^2$, i.e. 
  \begin{equation}
   \hat{\cF}^{-1} L^2_\pm(\realti)=H_\pm^2, \label{3.40a}
  \end{equation}
  $H_+^2$ denoting a Hardy space on the upper half-plane and 
  $H_-^2$ denoting a Hardy space on the lower half-plane. 
  If we apply the first of equations \eqref{3.33} 
  to the time factors \eqref{3.34a} and \eqref{3.ca}, we then have 
  \begin{align}
   \tilde{T}^\pm_n(E)=&\frac{A^\pm_n}{\sqrt{2\pi\hslash}}
   \int_{-\infty}^{\infty} 
   e^{i(E-V_0)t/\hslash}e^{\mp(n+1/2)\gamma t}
   \theta(\pm t)dt, \notag\\ 
   =&\frac{\pm i}{\sqrt{\pi}} 
   \frac{\sqrt{(n+1/2)\hslash\gamma}}
   {E-V_0\pm i(n+1/2)\hslash\gamma}\in H_\pm^2, \label{3.35}
  \end{align}
  with the help of \eqref{3.k}. 
  Thus $ \tilde{T}^+_n(E)$ has a simple pole at the point
  $$V_0-i(n+1/2)\hslash\gamma\in\compm, $$ 
  and $ \tilde{T}^-_n(E)$ has a simple pole at the symmetrical point 
  $$V_0+i(n+1/2)\hslash\gamma\in\compp$$
  with respect to the real axis. 
  To obtain the original equations \eqref{3.34a} we note that, 
  from an application of the second of equations \eqref{3.33} 
  to \eqref{3.35}, the integral being taken along 
  a large semicircle in the lower half-plane when $t>0$ and 
  in the upper half-plane when $t<0$. 
  
  By taking the squares of the moduli of \eqref{3.35} respectively, 
  we obtain 
  \begin{equation}
   \bigl| \tilde{T}^-_n(E)\bigr|^2 =
    \bigl| \tilde{T}^+_n(E)\bigr|^2 =
    \frac{1}{\pi}\frac{(n+1/2)\hslash\gamma}
    {(E-V_0)^2+[(n+1/2)\hslash\gamma]^2}. \label{3.36}
  \end{equation}
  These physical energy distributions are the 
  {\it Breit-Wigner resonance formulas}, 
  having the same resonance energy $V_0$ and the half-widths 
  \begin{equation}
   \Gammav^\pm_n=\left( n+\frac{1}{2}\right)\hslash\gamma. \label{3.50}
  \end{equation}
  Specially, the quantity $\Gammav^+_n$ is what is sometimes called 
  the {\it decay width}. 
  Equations \eqref{3.50} inform us that 
  {\it the  half-widths are quantized 
  in the unstable system with the Hamiltonian \eqref{3.1}}. 
  The integrals of \eqref{3.36} with respect to $E$ are 
  \begin{equation}
   \int_{-\infty}^{\infty}\bigl| \tilde{T}^-_n(E)\bigr|^2 dE=
    \int_{-\infty}^{\infty}\bigl| \tilde{T}^+_n(E)\bigr|^2 dE=1, 
    \label{3.61} 
  \end{equation}
  corresponding to \eqref{3.i}, 
  so {\it $\tilde{T}^\pm_n(E)$ are normalized}. 
  
  Let us take the products of 
  the mean lifetimes $(\Deltav t)^\pm_n$ and 
  the half-widths $\Gammav^\pm_n$. 
  From \eqref{3.52} and \eqref{3.50} these products read 
  \begin{equation}
   (\Deltav t)^\pm_n \Gammav^\pm_n
    =\frac{\hslash}{2}. \label{3.51}
  \end{equation}
  These results are independent of $n$. 
  {\it There are thus Heisenberg's uncertainty relations 
  between the mean lifetimes and the half-widths}. 
  
 \section{Discussion} \label{sect.4}
 We have obtained some insight into the main features 
 of the unstable system by making a study of 
 the one-dimensional parabolic potential barrier. 
 The results of solving 
 the time-independent Schr\"{o}dinger equation for the unstable system 
 are complex energy eigenvalues 
 and generalized eigenfunctions in Gel'fand triplet. 
 These generalized eigenfunctions are non-stationary states. 
 We have assumed that 
 the time factor corresponding to a non-stationary state 
 is a square-integrable function. 
 This assumption requires that 
 the states of complex energy eigenvalues in the lower half-plane 
 exist only on the future part 
 and those of complex energy eigenvalues in the upper half-plane
 exist only on the past part. 
 They correspond to the states of decay and growth, respectively. 
 The distributions of physical energy of 
 these decaying and growing states 
 are given by the Breit-Wigner resonance formulas 
 whose half-widths are determined by the complex energy eigenvalues. 
 Thus the half-widths are quantized in the unstable system 
 instead that the values of physical energy are quantized 
 in the stable system (e.g. the harmonic oscillator, the hydrogen atom). 
 These results are capable of comparison with experiments. 
 The observational results will provide a confirmation of 
 the complex energy eigenvalues and 
 generalized eigenfunctions in Gel'fand triplet. 
 
 We may discuss the comparison with 
 the formulation given by Bohm and Gadella~\cite{bohm}. 
 According to their formulation 
 Gamow vectors (i.e. the states of resonance 
 corresponding to a pair of simple poles of the $S$-matrix) 
 are represented by generalized functions 
 of the following Gel'fand triplets, 
 \begin{equation*}
  \begin{CD}
   \cS_\pm(\realti) @. \,\,\,\subset\,\,\, @. 
   L^2_\pm(\realti) @. \,\,\,\subset\,\,\, @. 
   {\cS_\pm(\realti)}^\times  \\
   @V{\hat{\cF}^{-1}}VV @. 
   @V{\hat{\cF}^{-1}}VV @. 
   @VV{\left(\hat{\cF}^{-1}\right)^\times}V \\
   \Deltav_\pm @. \subset @. 
   H_\pm^2 @. \subset @. 
   {\Deltav_\pm}^\times, \label{3.40}
  \end{CD}
 \end{equation*}
 where $\cS_\pm(\realti)\equiv\cS(\realti)\cap L^2_\pm(\realti)$ and 
 $\Deltav_\pm\equiv\cS(\realen)\cap H^2_\pm$. 
 The time evolution on ${\Deltav_+}^\times$ is well defined 
 only on the future part 
 and that on ${\Deltav_-}^\times$ is well defined 
 only on the past part. 
 We could, however, express the time factor 
 as a Hilbert vector (\S~\ref{sect.3}), while 
 we described the time-independent wave function 
 by Gel'fand triplets (\S~\ref{sect.2}). 
 We thus see that the origin of the separation 
 of the domain of the time evolution 
 lies in the property of the square-integrable of the time factor. 
 
 In \S\S~\ref{sect.3.1} and \ref{sect.3.2} we supposed that 
 there is no interference between the growth and decay processes. 
 In dealing with these processes, 
 a further possibility to be taken into consideration is 
 the direct scattering~\cite{dirac}. 
 In this case, we must first require 
 {\it the continuity condition for the time factor at time $t=0$} 
 \begin{equation}
  T^-_n(0)=T^+_n(0). \label{4.j} 
 \end{equation}
 The reason for this condition \eqref{4.j} is that 
 the time-dependent Schr\"{o}dinger equation \eqref{1.1} is 
 linear in the operator $\partial/\partial t$. 
 We can now require 
 {\it the normalizing condition for the time factor}, 
 instead of \eqref{3.i} 
 \begin{equation}
  \int_{-\infty}^\infty 
   \left|T^-_n(t) +T^+_n(t)\right|^2 dt =1. \label{4.i} 
 \end{equation} 
 This condition \eqref{4.i} means that 
 the total probability of the state $n$ being at any time is unity. 
 The total energy dependence, 
 namely $\tilde{T}^-_n(E)$ plus  $\tilde{T}^+_n(E)$, 
 is a real function of $E$. 
 By taking the square of the modulus of the sum, 
 we obtain 
 \begin{equation}
  \bigl| \tilde{T}^-_n(E) +\tilde{T}^+_n(E)\bigr|^2 
   =\frac{2}{\pi}\frac{[(n+1/2)\hslash\gamma]^3}
   {\{(E-V_0)^2+[(n+1/2)\hslash\gamma]^2\}^2} \label{4.36}
 \end{equation}
 Comparing this with \eqref{3.36}, 
 we see that the distribution \eqref{4.36} has 
 a narrow width (its quarter-width is equal to 
 $\Gammav^\pm_n$ given by \eqref{3.50}) and 
 a sharp peak (its maximum value is twice that of \eqref{3.36}) 
 at the same point $E=V_0$. 
 This distribution is different from 
 the Breit-Wigner resonance formula. 
 It is necessary to re-examine experimental data of 
 an absorption line induced by the scattering process. 
 We will investigate how this difference would appear experimentally. 
 
 The foregoing work shows how 
 the properties of the time factor 
 of the non-stationary state (of the unstable system) 
 are similar to the properties of the time-independent wave function 
 of the bound state (of the stable system), and vice versa. 
 Further, the physical energy of the unstable system is 
 the continuous spectrum 
 running from $-\infty$ to $\infty$, 
 and the lowest energy does not exist. 
 We should thus expect to find {\it time operators} 
 for the unstable system in quantum mechanics. 
 The subject for further investigation is 
 to get these time operators 
 and set up a relativistic quantum mechanics 
 treating the time and coordinates 
 {\it as observables} on the same footing. 
 
 \appendix
 \section*{Appendices}
 \section{The harmonic oscillator} \label{sect.a} 
 The Hamiltonian of the one-dimensional harmonic oscillator is
 \begin{equation}
  \hat{H} =-\frac{\hslash^2}{2m} \frac{d^2}{dx^2}
   +V_0 +\frac{1}{2} m\omega^2 x^2, \label{2.1}
 \end{equation}
 where $V_0\in\real$ is the minimum potential energy, 
 $m>0$ is the mass and $\omega>0$ is the angular frequency. 
 This Hamiltonian is an essentially self-adjoint on 
 a Schwartz space $\cS(\realco)$. 
 Further, two conditions are satisfied~\cite{bohm,bogolubov}. 
 \begin{enumerate}
  \item $\cS(\realco)$ is an invariant subspace of $\hat{H}$. 
  \item $\hat{H}$ is continuous on $\cS(\realco)$. 
 \end{enumerate}  
 
 The time-independent Schr\"{o}dinger equation \eqref{1.6} is
 \begin{equation}
  -\frac{\hslash^2}{2m} \frac{d^2}{dx^2}u(x)
   +\left( V_0 +\frac{1}{2} m\omega^2 x^2\right) u(x) 
   = E u(x). \label{2.5}
 \end{equation}
 It is convenient to introduce the dimensionless variables 
 \begin{align}
  \xi\equiv&\alpha x, \,\,\, 
  \alpha\equiv\sqrt{\frac{m\omega}{\hslash}}, \label{2.6}\\
  \lambda\equiv&\frac{2(E-V_0)}{\hslash\omega}.
 \end{align}
 Equation \eqref{2.5} now becomes 
 \begin{equation}
  \frac{d^2 u}{d\xi^2}
   +\left(\lambda -\xi^2\right)u=0. 
 \end{equation}
 Under the boundary condition $u\in\cS(\realco)$, 
 this equation can be solved by the method of 
 power series expansion.
 The energy eigenvalues are
 \begin{equation}
  E_n=V_0+\left(n+\frac{1}{2}\right)\hslash\omega
   \in\real \,\,\, 
   \left(n=0, 1, 2, \cdots\right) \label{2.7}
 \end{equation}
 and the eigenfunctions are
 \begin{align}
  u_n(x)=&N_n e^{-\alpha^2x^2/2} H_n(\alpha x) 
  \in\cS(\realco), \label{2.8}\\
  N_n=&\left(\frac{\alpha}{2^n n! \sqrt{\pi}}\right)^\frac{1}{2}. 
  \label{2.c}
 \end{align}
 Here $H_n(\alpha x)$ are the Hermite polynomials of degree n, and 
 the constants $N_n$ are determined by 
 the normalizing condition \eqref{1.8}. 
 The set of eigenfunctions 
 $\left\{u_n\right\}_{n=0}^{\infty}$, 
 given by \eqref{2.8}, 
 forms a complete orthonormal system on 
 a Lebesgue space $L^2(\realco)$~\cite{schwartz}. 
 
 The energy eigenvalues \eqref{2.7} show that
 the probability densities of 
 one-dimensional harmonic oscillator are time-independent, 
 i.e. the eigenfunctions \eqref{2.8} are stationary states.
 
 \section{Properties of the polynomials $H^\pm_n(\xi)$} \label{sect.b}
 The differential equations \eqref{3.15} with \eqref{3.25} 
 \begin{equation}
  \frac{d^2H^\pm_n}{d\xi^2} 
   \pm 2i\xi\frac{dH^\pm_n}{d\xi} 
   \mp 2in H^\pm_n=0, 
   \label{a.1}
 \end{equation}
 where $H^\pm_n(\xi)$ are the polynomials of degree $n$.
 If we define them so that 
 the highest power of $\xi$ appears with the coefficient $2^n$, 
 the first few polynomials calculated from \eqref{3.19} are 
 \begin{align*}
  H^\pm_0(\xi)=&1, \,\,\, H^\pm_1(\xi)=2\xi, \,\,\,
  H^\pm_2(\xi)=4\xi^2 \mp 2i, \\
  H^\pm_3(\xi)=&8\xi^3 \mp 12i\xi, \,\,\,
  H^\pm_4(\xi)=16\xi^4 \mp 48i\xi^2 -12. 
 \end{align*}
 These polynomials in general can be expressed as
 \begin{equation}
  H^\pm_n(\xi)=\left(\mp i\right)^n e^{\mp i\xi^2}
   \frac{d^n}{d\xi^n}e^{\pm i\xi^2}. 
   \label{a.2}
 \end{equation} 
 Thus the parity of $H^\pm_n(\xi)$ is
 \begin{equation}
  H^\pm_n(-\xi)=(-)^n H^\pm_n(\xi), 
   \label{a.b}
 \end{equation}
 and the conjugate complex of $H^\pm_n(\xi)$ is
 \begin{equation}
  H^\pm_n(\xi)^*=H^\mp_n(\xi). 
   \label{a.c}
 \end{equation}
 
 An alternative way of defining the polynomials 
 $H^\pm_n(\xi)$ is as the generating functions 
 $S^\pm(\xi, s)$ given by 
 \begin{equation}
  S^\pm(\xi, s)\equiv e^{\mp i(2s\xi -s^2)}
   =\sum_{n=0}^\infty\frac{H^\pm_n(\xi)}{n!}(\mp is)^n. 
   \label{a.3}
 \end{equation}
 From this definition we obtain the recurrence formulas 
 \begin{align}
  &H^\pm_{n+1}-2\xi H^\pm_n
  \pm 2in H^\pm_{n-1}=0, \label{a.4}\\
  &dH^\pm_n/d\xi=2n H^\pm_{n-1}. \label{a.5} 
 \end{align}
 Corresponding to the orthogonality relation of the Hermite polynomials, 
 we now have, from \eqref{a.3} 
 \begin{equation}
  \int_{-\infty}^{\infty}
   {H^\mp_m(\xi)}^* H^\pm_n(\xi) e^{\pm i\xi^2} d\xi 
   =(\pm 2i)^n n! \sqrt{\pm i\pi} \delta_{mn}, 
   \label{a.6}
 \end{equation}
 where $\delta_{mn}$ is the Kronecker delta symbol.

\end{document}